# Investigation of gating effect in Si spin MOSFET

Soobeom Lee[1], Fabien Rortais[1], Ryo Ohshima[1], Yuichiro Ando[1,*], Minori Goto[2], Shinji Miwa[2,$], Yoshishige Suzuki[2], Hayato Koike[3] and Masashi Shiraishi[1,#]

1. Department of Electronic Science and Engineering, Kyoto University, Kyoto 615-8510, Japan.
2. Graduate School of Engineering Science, Osaka University, Osaka 560-8531, Japan.
3. Advanced Products Development Center, TDK Corporation, Chiba 272-8558, Japan.

* Corresponding author: Yuichiro Ando (ando@kuee.kyoto-u.ac.jp)
# Corresponding author: Masashi Shiraishi (mshiraishi@kuee.kyoto-u.ac.jp)
$ Present address: Institute for Solid State Physics, The Univ. Tokyo, Kashiwa, Chiba 277-8581, Japan.

**Abstract**

A gate voltage application in a Si-based spin metal-oxide-semiconductor field-effect transistor (spin MOSFET) modulates spin accumulation voltages, where both electrical conductivity and drift velocity are modified while keeping constant electric current. An unprecedented reduction in the spin accumulation voltages in a Si spin MOSFET under negative gate voltage applications is observed in a high electric bias current regime. To support our claim, the electric bias current dependence of the spin accumulation voltage under the gate voltage applications is investigated in detail and compared to a spin drift diffusion model including the conductance mismatch effect. We proved that the drastic decrease of the mobility and spin lifetime in the Si channel is due to the optical phonon emission at the high electric bias current, which consequently reduced the spin accumulation voltage.

Semiconductor spintronics is an emerging research field that exploits the spin degree of freedom in semiconducting materials as a new paradigm electronics. Spin-dependent signals are electrically detected and modulated by external electric/magnetic fields in spin-driven semiconductor device, which permits nonvolatile and low-energy-consumption information processing.[1,2] Therefore, the semiconductor-based spintronic devices have attracted a lot of attention because of their multifunctionality. Si has good spin coherence due to its weak spin-orbit interaction[3–7] and good compatibility with large scale integration technologies, which have motivated many studies towards realizing potential of Si-based spintronic devices. At the first step, highly doped Si has been used for spin transport to sidestep the conductivity mismatch,[6,8,9] however, the screening effect due to ionized impurities hinders effective gating. Therefore, non-degenerate Si is preferable as a spin transport channel for practical applications towards semiconductor spin devices. Recently, operation of a spin metal-oxide-semiconductor field-effect transistor (spin MOSFET), which is a typical spin transistor, has been successfully demonstrated at room temperature,[10,11] where spin-dependent signals were modulated by electrical gating via a buried oxide layer. Despite the successful spin transistor operation at room temperature, the mechanism of gate-dependent spin accumulation signals in Si spin MOSFET is still elusive. Therefore, it is important to understand how spin signals are modulated by gating to improve Si spin MOSFET.

A schematic of a typical Si spin MOSFET is shown in Fig. 1(a). The Si spin MOSFET device consists of two Fe/Co ferromagnetic (FM) electrodes with MgO tunnel barrier and two non-magnetic electrodes on a lateral channel of the non-degenerate Si. For the channel, we implanted phosphorus (P), as an n-type dopants, into the top silicon layer of a silicon-on-insulator substrate with the following structure: 100-nm-thick Si(100)/200-nm-thick $SiO_2$/bulk Si(100). The dopant concentration of the channel was ≈$5 \times 10^{17}$ cm$^{-3}$ and was lower than the degenerate limit of $3.5 \times 10^{18}$ cm$^{-3}$ for n-type Si, indicating that the Si channel in this study is non-degenerate.[12] Highly doped silicon with a thickness of 20 nm was grown on the non-degenerate Si by magnetron sputtering to suppress the formation of a depletion layer.[13] MgO(0.8 nm)/Co(0.6 nm)/Fe(12.4 nm)/Au (3 nm) layers were subsequently deposited by molecular beam epitaxy for ferromagnetic contacts. After the deposition of the layered structure, two FM contacts was formed on the channel by using electron beam lithography and argon-ion milling. The dimensions of the FM1 and FM2 contacts were $0.2 \times 21$ μm$^2$ and $0.8 \times 21$ μm$^2$, respectively, to achieve different coercive fields. The edge-to-edge distance between two FM contacts was set to 1.3 μm. It should be noted that the top surface of the Si channel was over-etched to remove the highly doped Si layer on the non-degenerate Si channel. Finally, two non-magnetic (NM) contacts were fabricated outside the FM contacts as reference electrodes. Spin signals were measured at 300 K by using non-local four-terminal (NL4T)[14] and local three-terminal (L3T)[15] methods.

Figures 1(b) and (c) show spin signals measured in the NL4T and the L3T geometries, respectively. The observation of clear rectangular magnetoresistances (MR) ($\Delta V_{\text{NL4T}}$ and $\Delta V_{\text{L3T}}$) indicates successful spin transport in the non-degenerate Si channel in both the NL4T and the L3T geometries. The amplitudes of the MR in the NL4T and the L3T methods were approximately 56 μV and 1.35 mV under an injection current of 1 mA, respectively. $I$-$V$ curves as a function of the back-gate voltage ($V_G$) were measured using a conventional four terminal method to determine the conductivity of the Si channel (see Fig. 1(d) and its inset). The channel conductivity ($\sigma_{\text{Si}}$) was strongly modified by carrier accumulation/depletion by the gating. Figures 1(e) and (f) show $I$-$V$ curves of the two FM contacts under the application of $V_G$ from − 10 V to 40 V, i.e. the gate voltage dependence of the FM contact resistance. The $I$-$V$ curves are identical, which unequivocally rules out the modification of the interface resistance of the FM contacts under the gate voltage applications.

Next, we controlled $\Delta V_{\text{L3T}}$, an output voltage in the Si spin MOSFETs, by electrical gating. The electric bias current ($I_{\text{inj}}$) in the L3T geometry was set to be from 0.1 mA to 1.5 mA. Figure 2(a) shows the $V_G$ dependence of $\Delta V_{\text{L3T}}$ with various $I_{\text{inj}}$ values. $\Delta V_{\text{L3T}}$ was inversely dependent on $V_G$ in a low electric bias current region. However, in a high electric bias current region ($I_{\text{inj}} \geq 0.6$ mA), the signal amplitude was decreased under negative $V_G$ unlike in the low-bias regime. We replotted the results by interchanging the horizontal axis from $V_G$ to $\sigma_{Si}$ as shown in Fig. 2(b) for $I_{inj}$ values of 0.2 mA and 1.5 mA. At high injection current, the spin accumulation voltage is not proportional to $1/\sigma_{Si}$ unlike the previous report.[16] Specifically, it should be noted that a decrease in the signal amplitude under high $I_{inj}$ and negative $V_G$ regimes is unusual. Possible mechanisms to explain the $V_G$ dependence of $\Delta V_{L3T}$ are as follows: (i) Carriers are accumulated/depleted in the Si channel adjacent to the $SiO_2$ (gate insulator) layer by positive/negative gate electric field applications, which changes spin current path in the Si. Since spin scattering centers are at the interface of the Si and the $SiO_2$, which affects spin lifetime and/or the diffusion constant as proposed in the literature.[17] (ii) the conductivity of the Si channel is modified by the gating, which induces reappearance of the conductivity mismatch.[18] (iii) the electric field across the Si channel can give rise to modulation of various spin transport characteristics, such as drift velocity, mobility, diffusion constant and spin lifetime, resulting in the modulation of the spin signals.[19,20]

We now proceed to discuss spin scattering at the $Si/SiO_2$ interface under the gate voltage application. The Hanle effect in the NL4T geometry was measured with an injection current ($I_{inj}$) of 0.5 mA and by changing gate voltages to estimate the spin lifetime ($\tau_s$) and the diffusion constant ($D$) in the Si. The Hanle spin precession signals were observed as shown in Fig. 3(a) and were fitted by an analytical solution of the following function, where the widths of two FM contacts were considered:[21,22]

$$\frac{V_{Hanle}^{AP-P}}{I_{inj}} = S_0 \int_{-w_{FM}^{inj}}^{0} \int_{d}^{d+w_{FM}^{det}} \int_{0}^{\infty} \frac{1}{\sqrt{4\pi D t}} \exp\left(-\frac{(x_{inj}-x_{det})^2}{4Dt}\right) \exp\left(-\frac{t}{\tau_s}\right) \cos(\omega_L t)\, dt dx_{det} dx_{inj} \quad , (1)$$

where $S_0$ is the constant that determines the signal amplitude, $d$ is the edge-to-edge gap distance between the ferromagnetic electrodes, $w_{FM}^{(inj/det)}$ is the width of the injector/detector FM contact, $\omega_L = g\mu_B B/\hbar$ is the Larmor frequency, $g$ is the Landé g-factor for the electrons ($g = 2$ in this study), $\mu_B$ is the Bohr magneton, and $\hbar$ is the Dirac constant. The fitting reproduces the experimental result as shown in Fig. 3(a). Figure 3(b) shows the $V_G$ dependence of $\tau_s$ (the top panel) and $D$ (the bottom panel), respectively, where $\tau_s$ and $D$ are constant following the gate voltages. It was reported that the spin lifetime in a lateral spin valve consisting of intrinsic Si was suppressed by the application of strong positive gate voltage, resulting in an enhancement of the Si/SiO$_2$ interface scattering.[17] Nevertheless, our experimental results imply that the spin scattering at the Si/SiO$_2$ interface of the Si spin MOSFET was not significant and/or the applied gate voltage regions in the study were not sufficient to induce the enhancement of the interfacial spin scattering. Hence, the first scenario is undoubtedly ruled out.

In the previous reports, the electric bias current dependence of the local spin signals in the spin MOSFET was well reproduced by the a conventional spin drift diffusion model[18,20] which takes into account the bias-dependent interfacial resistance of FMs. The details of the spin drift diffusion model discussed in the literature[18,20] are given in the Supplementary Materials.[23] In our study, the effect of electrical gating on the bias dependence of $\Delta V_{L3T}$ should also be taken into account in the spin drift diffusion model. The top panel of Fig. 4(a) shows the $I_{inj}$ dependence of $\Delta V_{L3T}$, where the gate voltage was changed from − 10 V to 30 V. From 0 V to 30 V, $\Delta V_{L3T}$ increases monotonically but almost linear and saturating with $I_{inj}$ as clarified in the literature.[18,20] Since a constant electric current was injected, the drift velocity of the spin-polarized electrons was reduced with increasing $V_G$, i.e., increasing the conductivity of the Si. Consequently, the magnitude of $\Delta V_{L3T}$ decreased with increasing $V_G$ due to a reduction in the spin drift velocity. Nevertheless, surprisingly, the $\Delta V_{L3T}$ at a $V_G$ of − 10 V was reduced in the high electric current regime, whereas a negative gate voltage should enhance the spin drift effect. The abnormality could be explained by optical phonon emission under a high electric field. In fact, both $\mu$ and $D$ in the Si are no longer constant under a high electric field greater than 5×10$^3$ V/cm, it is resulting in an increase in the Einstein ratio, $D/\mu$, under a high electric field.[24] The Einstein ratio is described as:

$$\frac{D}{\mu} = \frac{\delta}{\tanh(\delta)} \cdot \frac{D_0}{\mu_0} , \qquad (2)$$

$$\delta = \frac{E}{E_C}\left(1 - \exp\left(-\frac{\hbar\omega_0}{k_B T} \cdot \frac{E_C}{E}\right)\right) , \qquad (3)$$

where $D_0$ ($\mu_0$) is the diffusion constant (carrier mobility) at a low electric field, $E$ is the electric field in the channel, $E_C$ is the critical electric field that defines the onset of non-Ohmic behavior for drift velocity, $\hbar\omega_0$ is the energy for optical phonon emission ($\hbar\omega_0$ = 63 meV for Si), $k_B$ is the Boltzmann constant and $T$ is the crystal temperature ($T$ = 300 K in this study). From the above equations, the enhancement of the Einstein ratio may be regarded as an increase in the electron temperature. The Einstein ratio is calculated to be two-fold larger (at 1.5 mA, 7×10$^4$ V/cm) than the magnitude at a low electric bias current (at 0.1 mA), where we assume the electric field dependence of the Einstein ratio is identical to that reported in the literature.[24] Upstream ($\lambda_u$) and downstream ($\lambda_d$) spin transport length scales are given by $\frac{1}{\lambda_{d(u)}} = -(+)\left(\frac{D}{\mu}\right)^{-1}\frac{E}{2} + \sqrt{\left\{\left(\frac{D}{\mu}\right)^{-1}\frac{E}{2}\right\}^2 + \left(\frac{1}{\lambda_s}\right)^2}$, where $E$ and $\lambda_S$ ( = $(D\tau_s)^{0.5}$) are the electric field and spin diffusion length in the Si channel, respectively.[19] The modification of the Einstein ratio directly modifies the downstream and upstream spin transport length scales in Si, which should be considered in the theoretical model. Furthermore, the spin diffusion length under the high electric field is suppressed because both $D$ and $\tau_s$ are suppressed due to the optical phonon emission. The bottom panel of Fig. 4(a) shows the calculated $I_{inj}$ dependence of $\Delta V_{L3T}$ for various gate voltages, where the modification of the Einstein ratio and so the modified spin transport lengths are taken into account. The model including the aforementioned high electric field effect on $D$, $\mu$ and $\tau_s$ qualitatively reproduces the experimental result at $V_G$ of −10 V. The enhancement of the Einstein ratio is enough to explain the origin of the $I_{inj}$ dependence of $\Delta V_{L3T}$ under the positive gate voltage regime. For negative gate voltage, the suppression of the spin diffusion length should also be considered, because the enhancement of the Einstein ratio only does not reproduce the experimental result at $V_G$ = − 10 V (ref. 23; see also Supplementary Materials).

To understand how the suppression of the spin diffusion length plays a significant role at a negative gate voltage, the electric field dependence of the normalized spin diffusion length is shown in Fig. 4(b), where the spin diffusion length is determined to allow tracing the experimental results at $V_G$= −10 V. As it can be seen, the spin diffusion length was drastically reduced above 10$^4$ V/cm. The critical electric field from literature (5 × 10$^3$ V/cm)[25] is in good agreement with our experimental value when the spin diffusion length starts decreasing for $V_G = -10$ V, as shown in Fig. 4(b). The other supporting argument for the suppression of spin diffusion length at high electric field is the observation of a negative differential spin lifetime in intrinsic Si under strong electric fields.[26] Although the appearance of the decrease in the spin lifetime in Si was observed under a lower electric field in intrinsic Si, the behavior is considerably similar to our work (the slight difference in the critical electric fields is might be due to the different doping concentrations in Si). We emphasize that the enhancement of the Einstein ratio and the suppression of spin diffusion length under a high electric is ascribed to optical phonon emission under high electric fields in the gate voltage dependence of the spin signals in Si spin MOSFETs.

In summary, we found unprecedented suppression of the spin signals in Si spin MOSFET under a simultaneous application of negative gate voltages and high electric bias currents. The suppression is ascribed to optical phonon emission due to the high electric field application in the Si channel. The model calculation considering the conductance mismatch and the enhancement of the Einstein ratio due to the optical phonon emission reproduced the experimental results. These findings can pave a way towards further progress in Si spin MOSFET technology.

See Supplementary Materials for the details of the spin drift diffusion model, including the interfacial conductance mismatch and the model fitting of the experimental results under the gate voltage applications without considering the suppression of the spin diffusion length in the Si under a high electric field application performed in it, i.e., the negative differential spin lifetime effect.

S.L. acknowledges support from by the Japan Society for the Promotion of Science (JSPS) Research Fellow Program (Grant No. 18J22869). A part of this study was supported by a Grant-in-Aid for Scientific Research from the Ministry of Education, Culture, Sports, Science and Technology (MEXT) of Japan, Grant- in-Aid for Scientific Research (S) "Semiconductor Spincurrentronics" No.

16H06330. The authors are grateful to Prof. Hanan Dery (Univ. Rochester, USA) for his fruitful discussion on the negative differential spin lifetime, and Dr. Sachin Gupta for his encouragement and numerous thoughtful discussions.

**Figures and figure captions**

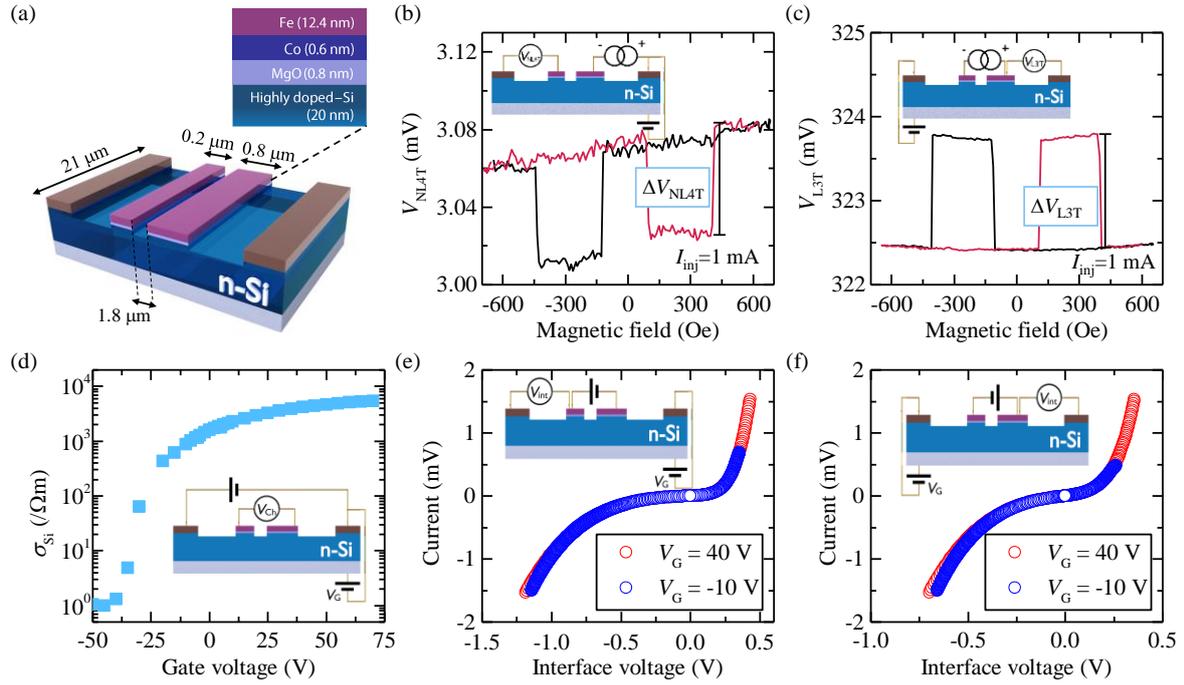

**Figure 1**

(a) Schematic of a Si spin MOSFET. The gate voltage is applied from the backside of the device via a gate oxide, $SiO_2$. (b) Typical magnetoresistnace measured in a non-local four-terminal (NL4T) geomtery. The measuring circuit is also shown in the figure. (c) Typical magnetoresistance measured in a local three-terminal (L3T) geometry. The measuring circuit is also shown in the figure. (d) Gate voltage dependence of the conductivity of the Si channel. The measuring circuit is also shown in the figure. *I-V* curves of (e) the FM1 contact and (f) the FM2 contact at gate voltages of − 10 V and + 40 V. The *I-V* curves are identical in all the ranges of the gate voltages (from − 10 V to + 40 V).

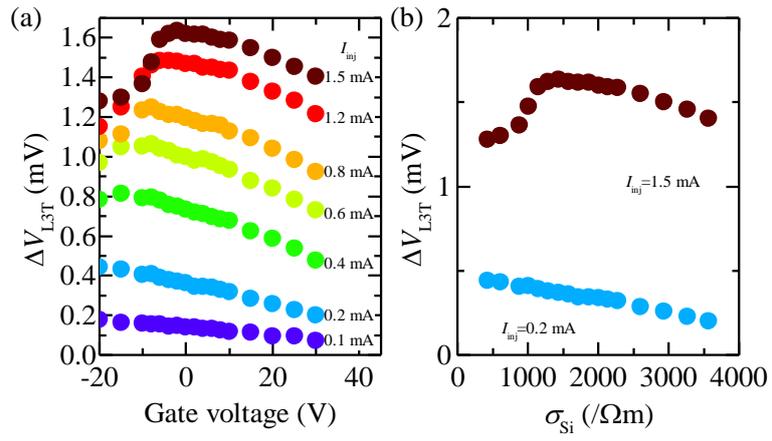

**Figure 2**

(a) Gate voltage dependence of spin accumulation voltages in the L3T geometry under various electric bias current applications. (b) Conductivity ($\sigma_{Si}$) dependence of spin accumulation voltages in the L3T geometry at $I_{inj}$ values of 0.2 mA and 1.5 mA.

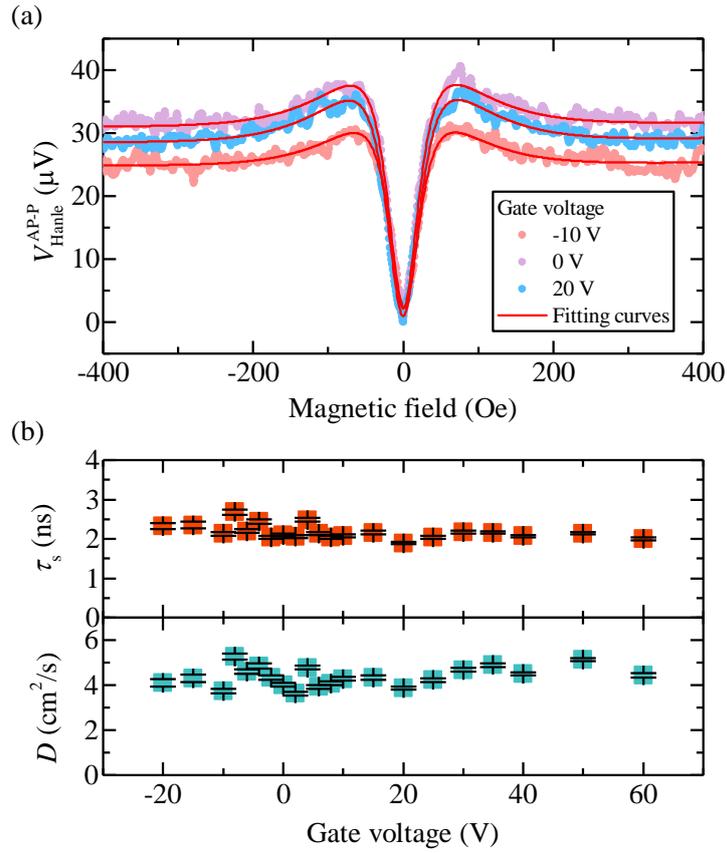

**Figure 3**

(a) Typical non-local Hanle-type spin precession signals under gate voltage applications. Closed circles are the experimental results. Solid lines are theoretical fitting lines obtained the function described in the main text. (b) Gate voltage dependence of the spin lifetime (the top panel) and the diffusion constant (the bottom panel). The error bars in the figure are the standard errors.

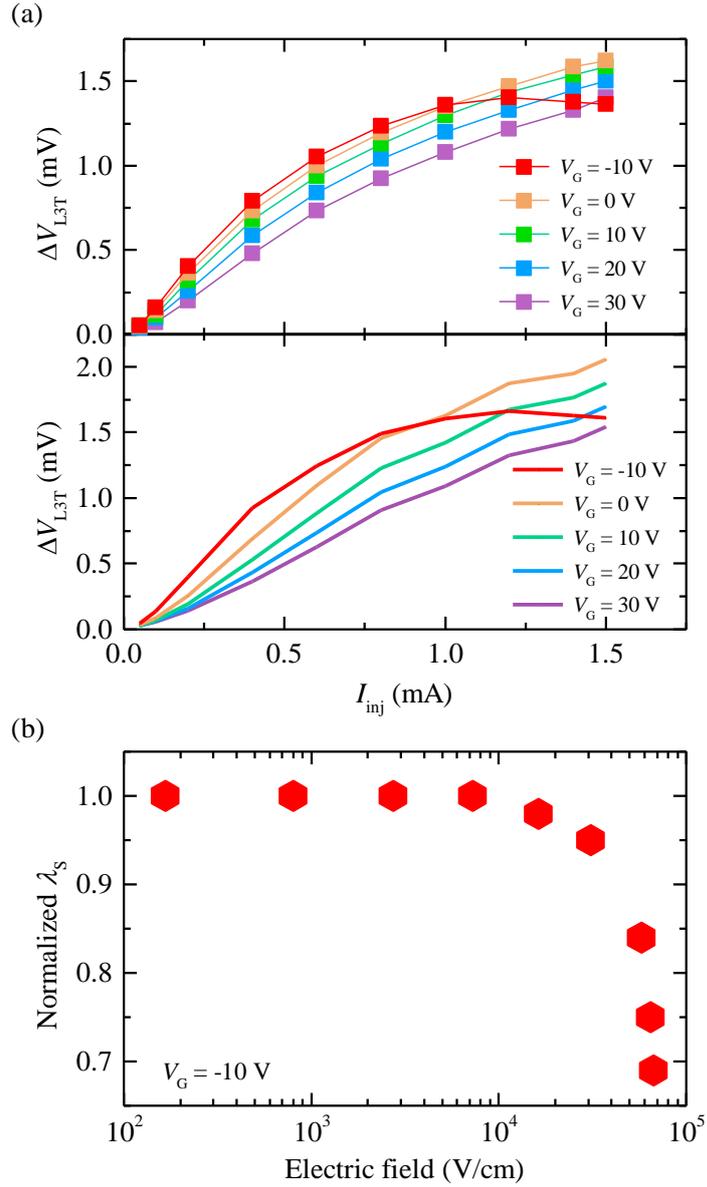

**Figure 4**

(a) Spin accumulation voltages as a function of the electric bias current and the gate voltage measured in the L3T scheme (the top panel) and spin accumulation voltages calculated using the spin drift diffusion model (the bottom panel). The details of the calculation are described in the main text. (b) Electric field dependence of normalized spin diffusion length. Spin diffusion length is determined to allow tracing the experimental results at a gate voltage of − 10 V.